\documentclass[iop,apj]{emulateapj}

\usepackage[USenglish]{babel}
\usepackage{amsmath,amssymb,amsxtra,amsfonts}
\usepackage{color}
\usepackage{graphicx}
\usepackage{hyperref}
\usepackage{natbib}
\newcommand{\lcdm}{\mathrm{\Lambda CDM}}
\newcommand{\densm}{\Omega_{\mathrm{m}}}
\newcommand{\densl}{\Omega_{\mathrm{\Lambda}}}
\newcommand{\densk}{\Omega_{\mathrm{k}}}
\newcommand{\hunit}{\mathrm{km \ s^{-1} \ Mpc^{-1}}}

\newcommand{\ud}{\mathrm{d}} %% For use between the integrand and the 
\newcommand{\der}[2]{\frac{\mathrm{d}#1}{\mathrm{d}#2}}

\shorttitle{Energy Condition Violation}
\shortauthors{C.-J.\ Wu et al.}

\begin{document}

\title{Reconstructing the History of Energy Condition Violation from 
Observational Data}

\author{Chao-Jian Wu\altaffilmark{1,2}}
\author{Cong Ma\altaffilmark{1}}
\author{Tong-Jie Zhang\altaffilmark{1,3}}\email{tjzhang@bnu.edu.cn}

\altaffiltext{1}{Department of Astronomy, Beijing Normal University,
Beijing, 100875, P.\ R.\ China}
\altaffiltext{2}{National Astronomical Observatories, Chinese Academy of 
Sciences, A20 Datun Road, Beijing 100012, P.\ R.\ China}
\altaffiltext{3}{Center for High Energy Physics, Peking University,
Beijing, 100871, P.\ R.\ China}

\begin{abstract}
    We study the likelihood of energy condition violations in the history of 
    the Universe.  Our method is based on a set of functions that characterize 
    energy condition violation.  FLRW cosmological models are built around 
    these ``indication functions''.  By computing the Fisher matrix of model 
    parameters using type Ia supernova and Hubble parameter data, we extract 
    the principal modes of these functions' redshift evolution.  These modes 
    allow us to obtain general reconstructions of energy condition violation 
    history independent of the dark energy model.  We find that the data 
    suggest a history of strong energy condition violation, but the null and 
    dominant energy conditions are likely to be fulfilled.  Implications for 
    dark energy models are discussed.
\end{abstract}

\keywords{cosmology: theory -- dark energy -- gravitation -- relativistic 
processes}

\section{Introduction}
\label{s:intro}

The {\em energy conditions} introduced by \citet[][Chapter 4]{1973CUPress} are 
coordinate-invariant inequality constraints on the energy-momentum tensor that 
appears in the Einstein field equation.  They play an essential role in the 
proof of theorems in the general theory of relativity concerning the existence 
of spacetime singularities \citep[see][]{1984UCPress}.  Due to their simplicity 
and model-independence, the energy conditions are listed as one of many 
approaches to understand the evolution of universe.

Of the many proposed energy conditions, the ones we analyze in this paper are 
the null, the strong, and the dominant energy conditions (abbreviated 
respectively as NEC, SEC, and DEC).  These energy conditions can be expressed 
as follows \citep[][Chapter 4]{2004NY} for a general stress-energy tensor $T$:
\begin{itemize}
    \item NEC: $T_{\alpha \beta} n^{\alpha} n^{\beta} \ge 0$ for all null 
	vectors $n$;
    \item DEC: $T_{\alpha \beta} t^{\alpha} t^{\beta} \ge 0$ and $T_{\alpha 
	\beta} T^{\beta}_{\gamma} t^{\alpha} t^{\gamma} \le 0$ for all timelike 
	vectors $t$;
    \item SEC: $T_{\alpha \beta} t^{\alpha} t^{\beta} \ge \frac{1}{2} 
	T^{\gamma}_{\gamma} t^{\delta} t_{\delta}$ for all timelike vectors 
	$t$.
\end{itemize}
The energy conditions are of great relevancy to the study of cosmology.  For
example, the NEC is an important condition of stability for fluids
\citep{2006JHEP...03..025D,2006PhRvD..74f3518B}, hence it is useful for the 
analysis of cosmological models.  The DEC guarantees stability of a source 
obeying it and imposes on the dark energy equation of state parameter $w$ a 
lower bound $w \ge -1$ \citep{2003PhRvD..68b3509C}.  Violation of the $w \ge 
-1$ bound can lead to the ``big rip doomsday'' scenario of the Universe 
\citep{2003PhRvL..91g1301C}.  It can also be shown that sudden future 
singularity solutions of the Friedmann-Lema{\^i}tre-Robertson-Walker (FLRW) 
cosmology always violate the DEC \citep{2004CQGra..21L.129L}.  Finally, SEC 
violation is a typical trait of a positive cosmological constant $\Lambda$ 
\citep[for example, see][]{2011CoTPh..56..525L} and other dark energy models 
\citep[see][and references therein]{2003A&A...402...53S}.  We therefore find an 
investigation of these energy conditions useful for our understanding of the 
Universe's evolution.

The knowledge of cosmic energy condition violation has been greatly facilitated 
by precise observational data.  \citet{1997Sci...276...88V,1997PhRvD..56.7578V} 
shows that Hubble constant and stellar age measurements, which bound the age of 
the Universe, suggest a history of SEC violation \citep[see also the review 
by][]{2003Sci...299...65K}.  X-ray galaxy cluster number count and type Ia 
supernova (SNIa) luminosity distance data have been applied to a constant-$w$ 
model, which also suggest SEC violation without significant indication for NEC 
violation \citep{2003A&A...402...53S}.  Similar results are also obtained by 
\citet{2008PhRvD..77h3518L} using available SNIa data at that time.  On the 
other hand, if one assumes the energy conditions, constraints on a variety of 
cosmological observables or parameters can be predicted, such as the Hubble 
parameter, luminosity and angular diameter distances, lookback time, total 
density parameter $\Omega(z)$, energy density $\rho(z)$, and pressure $p(z)$ 
\citep{2008CQGra..25p5013C}.

In this paper we study the issue of energy condition violations from a 
data-driven perspective.  Specifically, we aim at reconstructing what we can 
say about the likelihood of energy condition fulfillment (or violation) given 
the data.  To achieve this, we use the energy conditions themselves to 
construct a family of descriptive cosmological models for the recent history of 
cosmic expansion, and subject them to statistical test.  Throughout our paper 
we assume a universe of perfect fluids that lead to the spacetime solution 
characterized by the FLRW metric.  We use the Union2 SNIa luminosity distance 
data compilation \citep{2010ApJ...716..712A} and the observational Hubble 
parameter data from differential ages of passively evolving galaxies 
\citep{2005PhRvD..71l3001S,2010JCAP...02..008S} and radial baryon acoustic 
oscillation (BAO) measurements \citep{2009MNRAS.399.1663G}. The cosmological 
parameters used in this paper: $\densm=0.274$, $\densl=0.726$, and Hubble 
constant $H_0 = 74.2$ $\hunit$

This paper is organized as follows.  In Section \ref{s:ec} we specify our 
models built from our {\em indication functions} of energy condition violation 
and provide solutions for the cosmic expansion rate.  We proceed to Section 
\ref{s:analysis} where we lay out the procedures of analysis using luminosity 
distance and Hubble parameter data.  Our main results are presented in Section 
\ref{s:results} and are discussed in Section \ref{s:conc}.

\section{Energy Conditions and Cosmological Models}
\label{s:ec}

\subsection{Overview}
\label{s:ec:overview}

As summarized by \citet[][Chapter 4]{2004NY}, the energy conditions can be 
expressed by simple inequalities in terms of the energy density $\rho$ and the 
pressure $p$ in the setting of a homogeneous and isotropic FLRW universe:
\begin{itemize}
    \item NEC: $\rho + p \geq 0$,
    \item DEC: $\rho \geq 0$ and $-\rho \leq p \leq \rho$,
    \item SEC: $\rho + 3p \geq 0$ and $\rho + p \geq 0$.
\end{itemize}

By virtue of the Friedmann equation, the energy condition bounds can be 
expressed in terms of the Hubble constant-normalized, dimensionless expansion 
rate $E(z) = H(z) / H_0$ and its first derivative, or the deceleration 
parameter defined by
\begin{equation}
    q(z) = \frac{1 + z}{E(z)} \der{E}{z} - 1.
\end{equation}
The results, following \citet{2008PhRvD..77h3518L}, are given below:
\begin{itemize}
    \item NEC:
	\begin{eqnarray}
	    \label{ineq:nec}
	    q(z) - \frac{\densk (1+z)^2}{E^2(z)} + 1 \geq 0,
	  \end{eqnarray}
      
    \item DEC:
	\begin{eqnarray}
	    \label{ineq:dec}
	    2 - q(z) - \frac{2 \densk (1+z)^2}{E^2(z)} \geq 0,
	\end{eqnarray}
    \item SEC:
	\begin{eqnarray}
	    \label{ineq:sec}
	    q(z) \geq 0.
	\end{eqnarray}
      
\end{itemize}
A notable feature of these bounds is that they are only explicitly dependent on 
one arbitrary constant, namely the curvature parameter $\densk$, although the 
matter content of the FLRW universe does control the functional form of $E(z)$. 
However, for our purpose we do not need to make assumptions on the constituent 
matter of the universe being modeled.

\subsection{Models and Their Solutions for $E(z)$}
\label{s:ec:model}

We are therefore motivated to introduce the ``indication functions'' $F(z)$ 
that quantifies whether an energy condition has been violated.  Fulfillment (or 
violation) of an energy condition should be indicated by $F(z) \geq 0$ (or 
$F(z) < 0$).  The simplest choices are just the left-hand sides of inequalities 
(\ref{ineq:nec}-\ref{ineq:sec}).  We denote them as $F_{\mathrm{NEC}}, 
F_{\mathrm{DEC}}, \text{and } F_{\mathrm{SEC}}$ respectively:
\begin{itemize}
    \item NEC:
	\begin{eqnarray}
	    \label{eq:nec}
	    F_{\mathrm{NEC}}(z) = q(z) - \frac{\densk (1+z)^2}{E^2(z)} + 1,
	\end{eqnarray}
    \item DEC:
	\begin{eqnarray}
	    \label{eq:dec}
	    F_{\mathrm{DEC}}(z) = 2 - q(z) - \frac{2 \densk (1+z)^2}{E^2(z)},
	\end{eqnarray}
    \item SEC:
	\begin{eqnarray}
	    \label{eq:sec}
	    F_{\mathrm{SEC}}(z) = q(z).
	\end{eqnarray}

\end{itemize}

With the introduction of these indication functions, inequalities 
(\ref{ineq:nec}-\ref{ineq:sec}) become ordinary differential equations that can 
be integrated to find $E(z)$.  We may look at them as mappings from the set of 
dimensionless Hubble expansion rates $E(z)$, to that of indication functions 
$F(z)$, and vice versa.  These mappings enable us to reconstruct $F(z)$ from 
observables related to $E(z)$.

Our formalism so far remains general enough for any model that makes a 
prediction on $E(z)$.  In particular, we will concentrate on the underlying 
evolution of $E(z)$ that is favored by observational data\footnote{The practice 
of deriving a model of cosmic expansion by phenomenologically treating the 
model parameters without physical prescriptions is widely known as the 
``model-independent'' approach, which does not actually exclude the use of a 
model.}.  This approach has been used on the study of possible evolution of 
dark energy \citep{2003PhRvL..90c1301H,2005PhRvD..71b3506H} and the 
reconstruction of historical deceleration parameter $q(z)$ 
\citep{2006ApJ...649..563S}.

Following this line of inquiry, we can use a similar method to reconstruct the 
indication functions $F(z)$, where the function to be reconstructed is 
``coarse-grained'' as piecewise-constant.  While this can be done using more 
elaborate functions such as the smooth, hyperbolic tangent function 
\citep{2009JCAP...12..025C,2010PhRvD..81d3518Z}, we find the piecewise-constant 
functions greatly simplify the integration of equations 
(\ref{eq:nec}-\ref{eq:sec}).  In the interval in which $F(z)$ is constant, 
$F(z) = r$, the general solutions are given by
\begin{itemize}
    \item NEC:
	\begin{eqnarray}
	    E_{\mathrm{NEC}}(z) = \sqrt{C (1 + z)^{2r} - \frac{\densk (1 + 
	    z)^2}{r - 1}},
	\end{eqnarray}
    \item DEC:
	\begin{eqnarray}
	    E_{\mathrm{DEC}}(z) = \sqrt{C (1 + z)^{2 (3 - r)} + \frac{2 \densk 
	    (1 + z)^2}{2 - r}},
	\end{eqnarray}
    \item SEC:
	\begin{eqnarray}
	    E_{\mathrm{SEC}}(z) = \sqrt{C (1 + z)^{2 (r + 1)}},
	\end{eqnarray}
\end{itemize}
where $C$ stands for the integration constant.  To obtain the special solutions 
of $E(z)$ for the piecewise-constant $F(z)$ models extending from $z = 0$ to an 
arbitrary redshift, we can recursively apply the above solutions and the 
continuity condition of $E(z)$ across subinterval endpoints starting with $E(z 
= 0) = 1$, thereby fixing the integration constants.

\section{Analysis}
\label{s:analysis}

As we have briefly stated in the preceding section, we use the Fisher matrix of 
the full array of model parameters to extract a small number of principal 
components that preserve the information contained in the data without 
introducing serious over-parameterization.  The Fisher matrix elements $F_{ij}$ 
for the model parameters (not to be confused with the symbol for the indication 
functions $F$) are expressed by
\begin{equation}
    \label{eq:deffmatrx}
    F_{ij} = \langle - \frac{\partial^2 \ln P}{\partial \theta_i \partial 
    \theta_j} \rangle,
\end{equation}
where $P$ is the posterior probability density function (PDF) in the parameter 
space and $\theta_i$ is the $i$th model parameter, and the angled brackets 
stand for statistical averaging.  The principal components corresponds to the 
eigenvectors of the Fisher matrix.  The posterior probability $P$ is to be 
found from observational data.

\subsection{Posterior Probability}
\label{s:analysis:post}

To construct the posterior probability $P$ from observational data, we assume 
that the uncertainties assigned to the measurement results are Gaussian.  Under 
this assumption, the posterior probability can be expressed by an additive 
$\chi^2$ statistic with $\chi^2 = -2 \ln P$.  The generic form of $\chi^2$ 
under this assumption can be laid out as follows:
\begin{equation}
    \label{eq:chisqgeneric}
    \chi^2 = \sum_i \frac{(X^{\mathrm{th}}_i - 
    X^{\mathrm{obs}}_i)^2}{\sigma^2_i},
\end{equation}
where the symbols $X^{\mathrm{th,obs}}$ denotes, respectively, the theoretical 
or observed value of the $i$th observable (direct or indirect), and $\sigma_i$ 
is the $i$th Gaussian variance or uncertainty given by the data.  The summation 
is done over all individual data entries.

In dealing with parameterized models, it is natural to introduce nuisance 
parameters to be eliminated later by marginalization.  The nuisance parameters 
and their marginalization distorts the simple form of equation 
(\ref{eq:chisqgeneric}) as well as introducing computational complexities that 
often calls for Monte-Carlo techniques.  Nevertheless, before engaging any 
numerical computations we can eliminate certain nuisance parameters by 
analytical marginalization, as illustrated in the following sections.

\subsubsection{Luminosity Distance}
\label{s:analysis:post:lumin}

The luminosity distance data, such as the Union2 dataset that we use in this 
paper, are usually presented as tabulated distance moduli with errors.  
Physically, the luminosity distance modulus is the difference of apparent 
magnitude $m$ and the absolute magnitude $M$,
\begin{equation}
    \label{eq:defdistmod}
    \mu(z) = m - M = -5 + 5 \lg d_L(z),
\end{equation}
with $d_L$ measured in the units of $10$ parsecs.  It can be related to the 
modeled expansion rate of the universe, $E(z)$, by the formula
\begin{equation}
    \label{eq:lumindistdef}
    d_L(z) = \frac{c}{H_0} \frac{1 + z}{\sqrt{\left|\densk\right|}} 
    \operatorname{sinn} \left[\sqrt{\left|\densk\right|} \int^z_0 \frac{\ud 
    z'}{E(z')} \right],
\end{equation}
where the $\operatorname{sinn}$ function is a shorthand for the definition
\begin{eqnarray}
    \operatorname{sinn}(x) =
    \begin{cases}
	\sinh x, & \densk > 0, \\
	x, & \densk = 0, \\
	\sin x, & \densk < 0.
   \end{cases}
\end{eqnarray}

For notational convenience we employ a variable
\begin{equation}
    \label{eq:defmtilde}
    \tilde{m}(z) = 5 \lg \left[\frac{1 + z}{\sqrt{\left\vert \densk 
    \right\vert}} \operatorname{sinn} \left(
    \sqrt{\left\vert \Omega_k \right\vert} \int^z_0 \frac{\ud z'}{E(z')}
    \right)\right],
\end{equation}
where the parametric dependency on the Hubble constant $H_0$ has been separated 
out, leaving a dimensionless quantity as our model prediction.  With this 
definition the modeled distance modulus can be expressed as
\begin{equation}
    \label{eq:modeldistmod}
    \mu^{\mathrm{th}}(z) = \tilde{m}(z) + 5 \lg H_0 + M_0.
\end{equation}
In this expression we introduce a numerical constant $M_0$ to accommodate the 
numerical constants arising from unit conversions, as well as an additive 
variable characterizing the uncertain variability or spread of the standard 
candles' absolute magnitudes.  By doing so, we include $M_0$ as another 
nuisance parameter which affects the overall scaling of the modeled 
cosmological distance.

The $\chi^2$ statistic for a luminosity distance modulus dataset, according to 
equation (\ref{eq:chisqgeneric}), is therefore
\begin{equation}
    \label{eq:chisq:lumindist}
    \chi^2 = \sum_i \frac{\left[\tilde{m}(z_i) + 5 \lg H_0 + M_0 - 
    \mu^{\mathrm{obs}}(z_i)\right]^2}{\sigma_i^2}.
\end{equation}
The above expression is quadratic in the combination of nuisance parameters $(5 
\lg H_0 + M_0)$:
\begin{eqnarray}
    \label{eq:chisq:lumindist:quadratic:nuisance}
    \chi^2 = A(5 \lg H_0 + M_0)^2 + 2B (5 \lg H_0 + M_0) + C,
\end{eqnarray}
where, for notational simplicity, we have defined the symbols
\begin{equation}
    \label{eq:defabc}
    \begin{split}
	A &= \sum_i \frac{1}{\sigma_i^2}, \\
	B &= \sum_i \frac{\tilde{m}(z_i) - 
	\mu^{\mathrm{obs}}(z_i)}{\sigma_i^2}, \\
	C &= \sum_i \frac{\left[\tilde{m}(z_i) - 
	\mu^{\mathrm{obs}}(z_i)\right]^2}{\sigma_i^2}.
    \end{split}
\end{equation}

The nuisance parameters $H_0$ and $M_0$ can now be marginalized over by 
Gaussian integration over $(-\infty, \infty)$, leaving a ``reduced $\chi^2$'' 
in the form of
\begin{equation}
    \label{eq:chisq:lumindist:marginalized}
    \tilde{\chi}^2 = C- \frac{B^2}{A}.
\end{equation}
This expression of $\tilde{\chi}^2$ is a general result for luminosity distance 
data \citep{2003MNRAS.341.1299D,2004PhRvD..70d3531N}.

\subsubsection{Hubble Parameter}
\label{s:analysis:post:hub}

The observational Hubble parameter data we use are tabulated in Refs.\ 
\citep{2010AdAst2010E..81Z,2011ApJ...730...74M} and form an independent 
dataset.  As noted by \citet{2008JCAP...03..007Z}, the Hubble parameter data 
could help us catch possible ``wiggles'' of $E(z)$ predicted by certain dark 
energy models that would have been flattened out by the luminosity distance 
test.  For the Hubble parameter data our model requires equation 
(\ref{eq:chisqgeneric}) be in the form of
\begin{equation}
    \chi^2 = \sum_i \frac{\left[H_0 E(z_i) - 
    H^{\mathrm{obs}}(z_i)\right]^2}{\sigma_i^2}.
\end{equation}
We adopt the result of $H_0 = 74.2 \pm 3.6$ $\hunit$ measured from nearby 
Cepheids by the {\em Hubble Space Telescope} \citep{2009ApJ...699..539R} as a 
Gaussian prior, and use the analytic expression given by 
\citet{2011ApJ...730...74M} to construct the $\tilde{\chi}^2$ with $H_0$ 
marginalized over.  This statistic from $H(z)$ data is added to the one from 
SNIa data (eq.\ [\ref{eq:chisq:lumindist:marginalized}]) to give the final 
posterior log-PDF.

\subsection{Redshift Binning}
\label{s:analysis:binning}

\begin{figure}
    \includegraphics[width=.45\textwidth]{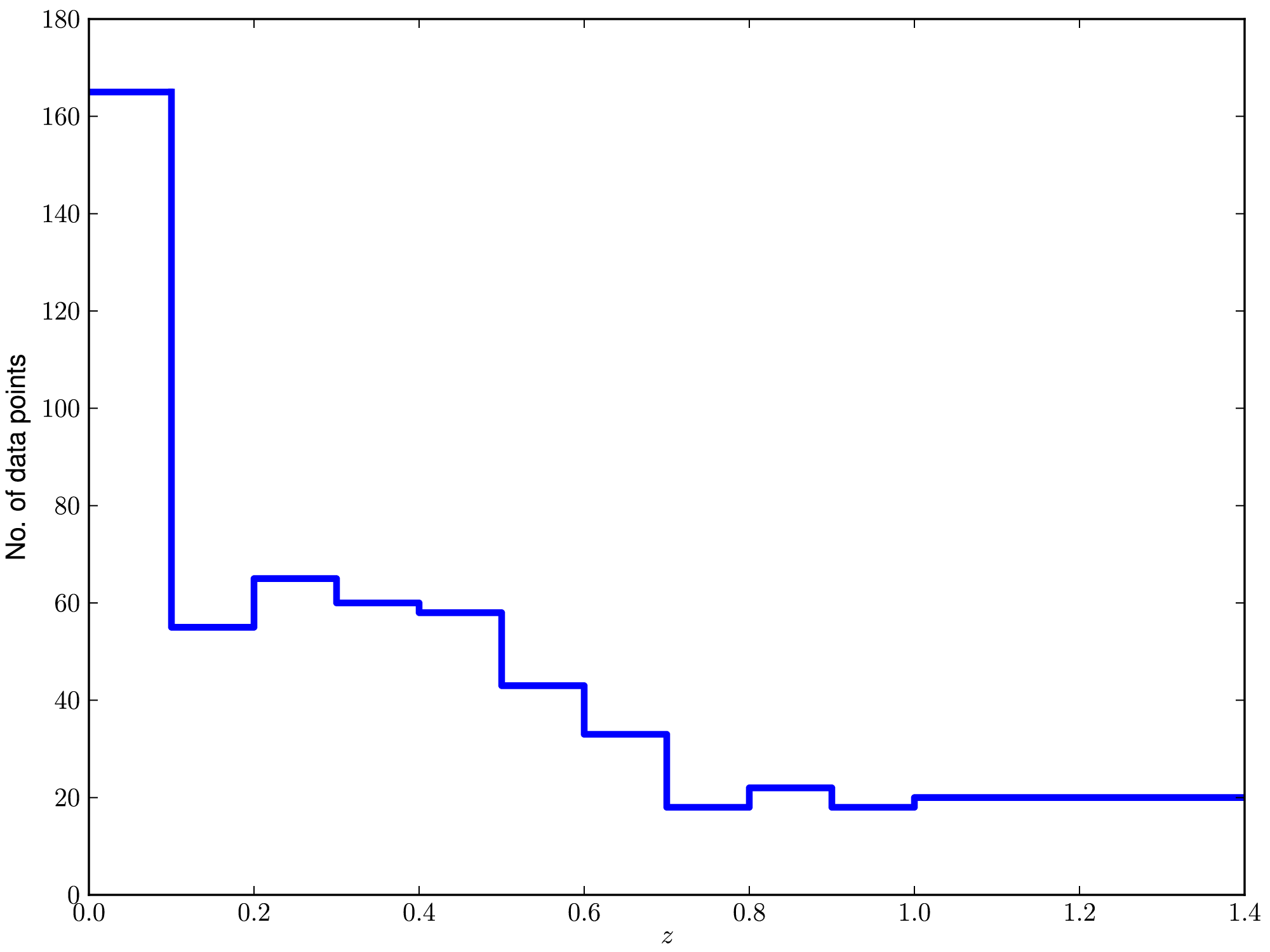}
    \caption{Binning of the redshift range to accommodate enough data in each 
    of the bins.  The histogram shows the number of SNIa data points 
    distributed across the redshift range, with each bin covering width 
    0.1.\label{fig:binning}}
\end{figure}

Ideally, the coarse-graining approximation of the $F(z)$ evolution should 
approach continuity as its limit.  However, for our analysis using the data of 
557 SNeIa from Union2 and the 13 $H(z)$ measurements, the redshift distribution 
of data poses a limit on how fine we can dissect the redshift range into bins, 
hence the resolution of our reconstruction of $F(z)$.  Each bin should cover 
enough data to make the value of $F(z)$ well-constrained inside of it.  To this 
end, we introduce 11 bins equally spaced to cover the redshift range, shown in 
Figure \ref{fig:binning}.  The first 10 bins are kept to be equal in width, but 
the last one is widened to cover the diminishing tail of the SNIa redshift 
distribution.

\subsection{Evaluation of the Fisher Matrix}
\label{s:analysis:fisher}

The Fisher matrix elements as defined by equation (\ref{eq:deffmatrx}) are 
statistical averages involving unknown parameters.  However, for our purpose we 
cannot evaluate them by averaging in the absence of a {\em known} posterior.  
In fact, it is neither practical nor logical to do this, as our final models to 
assess the evaluation of energy conditions are not the model described so far 
{\em per se}, but the models to be built from the principal components 
extracted from the Fisher matrix.

In practice, the Fisher matrix elements can be estimated at point in the 
parameter space where the posterior PDF is close to the maximum.  In other 
words, we can approximate the numerical values of the Fisher matrix elements 
using a {\em fiducial} model, which we choose to be the concordance flat 
$\lcdm$ model with $\densm = 0.274$, $\densl = 0.726$ as favored by combined 
observational data \citep{2010ApJ...716..712A,2011ApJS..192...18K} and $H_0 = 
74.2$ $\hunit$ from \citep{2009ApJ...699..539R}.  In a similar setting, 
\citet{2006ApJ...649..563S} used a even simpler, constant-input model as the 
fiducial one in their analysis, and found the method to be robust against the 
introducing of a simplified fiducial.  Our own findings are in agreement to 
their claim.

To add robustness and safeguard against limited sampling in the parameter 
space, we resampling \citep{1982jbor.book.....E} to check for any possible bias 
resulting from not treating the averaging in equation (\ref{eq:deffmatrx}) 
rigorously.  The result confirms the robustness of Fisher matrix estimation, 
and the average from the resampled Fisher matrices are used as the ``reference 
stack'' in subsequent steps.

The Fisher matrix elements are thus calculated using the $\tilde{\chi}^2$ (or 
posterior log-PDF) expressions given in the preceding Sections 
\ref{s:analysis:post:lumin} and \ref{s:analysis:post:hub}.  Notice that the 
normalization constants discarded in the previous steps do not contribute to 
the matrix elements because they are additive constants in the log-PDF 
independent of the parameters, and their derivatives vanish.

\section{Results}
\label{s:results}

\begin{figure}
    \includegraphics[width=.45\textwidth]{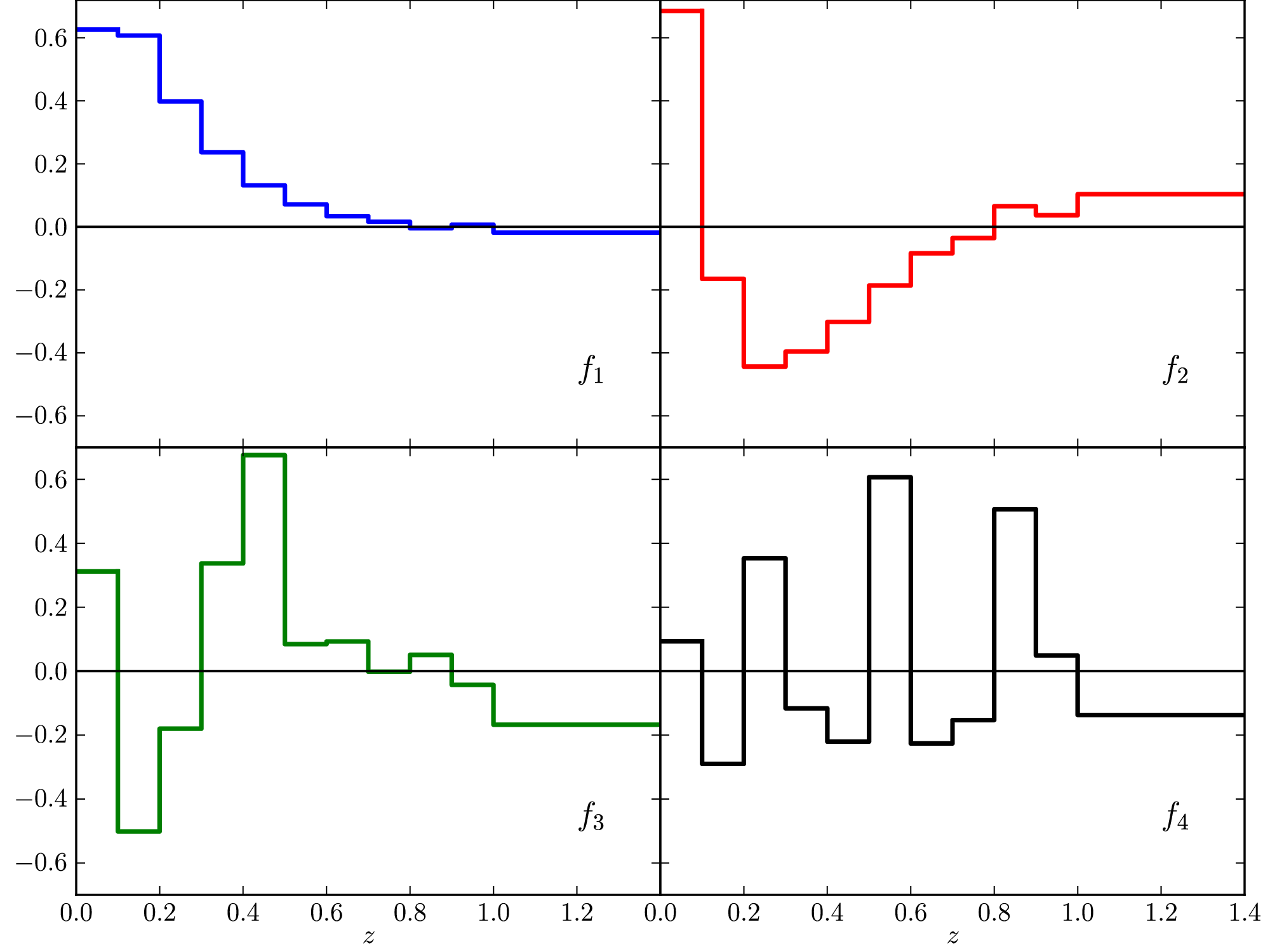}
    \caption{Four well-constrained principal components of 
    $F_{\mathrm{NEC}}(z)$.  They are normalized so that $\int{f^2_i(z) \ud 
    z=1}$.  $f_i$'s $(i = 1, 2, 3, 4)$ are the first four eigenvectors of the 
    Fisher matrix.\label{fig:modesNEC}}
\end{figure}

\begin{figure}
    \includegraphics[width=.45\textwidth]{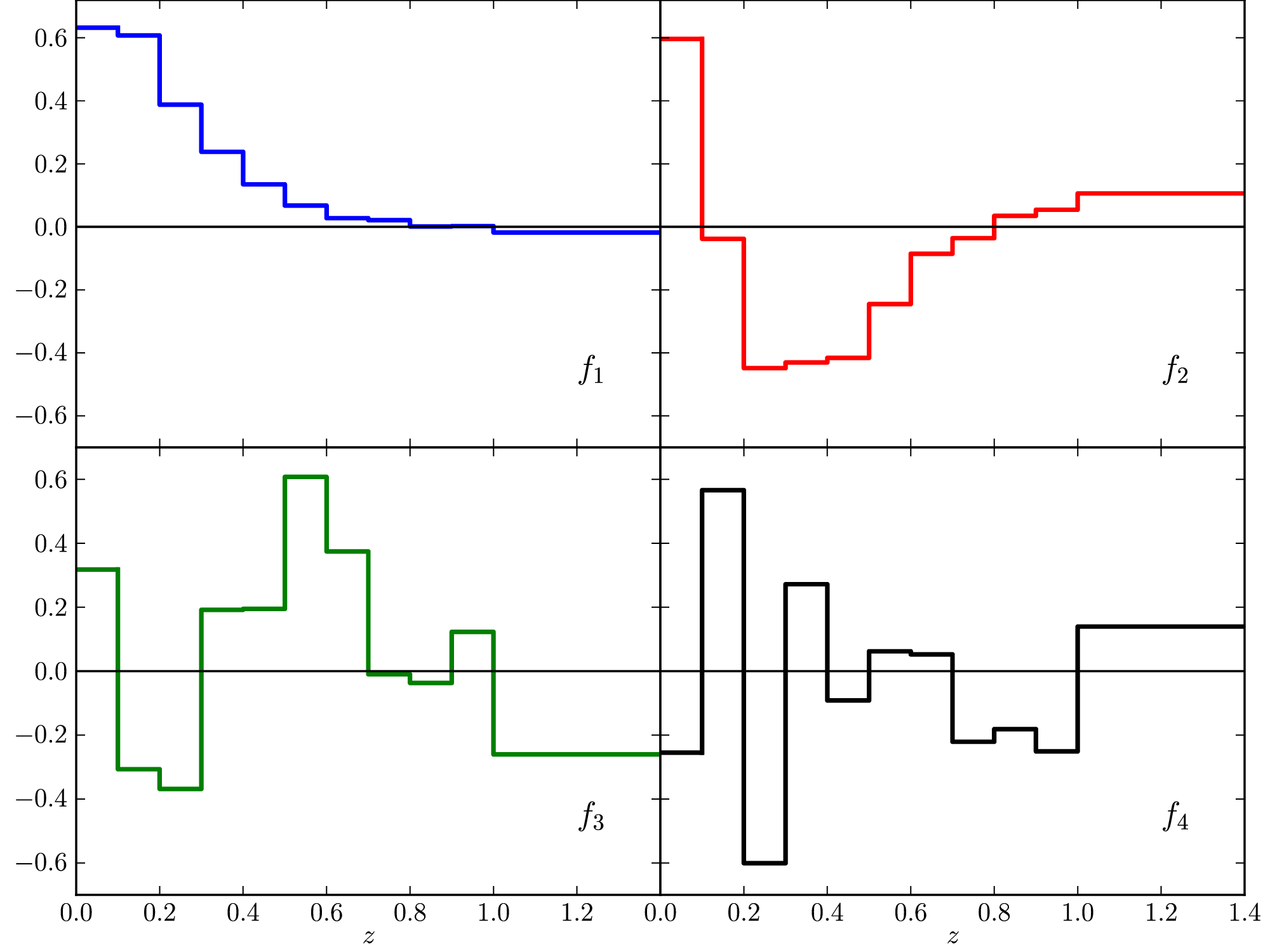}
    \caption{Same as Figure \ref{fig:modesNEC} but for 
    $F_{\mathrm{DEC}}(z)$.\label{fig:modesDEC}}
\end{figure}

\begin{figure}
    \includegraphics[width=.45\textwidth]{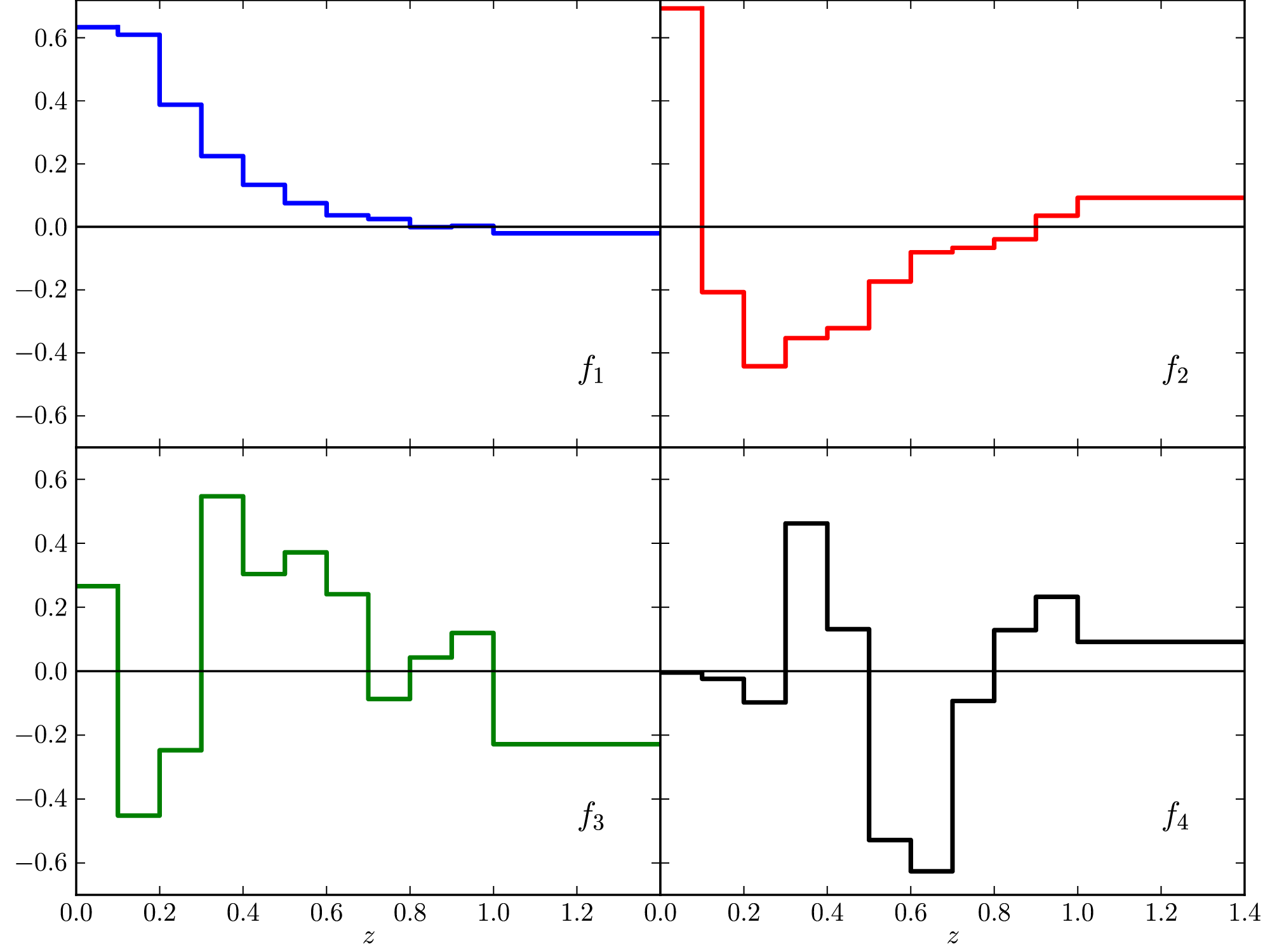}
    \caption{Same as Figure \ref{fig:modesNEC} but for 
    $F_{\mathrm{SEC}}(z)$.\label{fig:modesSEC}}
\end{figure}

We numerically compute the $(N + 1) \times (N + 1)$ Fisher matrix for $F(z)$ in 
each of the $N$ bins and $\densk$.  At this stage, the $\densk$ parameter can 
be effectively marginalized over by projecting the matrix onto the $N \times N$ 
subspace \citep[][Chapter 11]{2003moco.book.....D}.  The eigenvectors of the 
resulting matrix, $f_i$, are then retrieved.  In this manner, we can thus build 
a chain of models with successively more parameters.  The eigenvectors are the 
principal components we are after, and they are shown in Figures 
\ref{fig:modesNEC}, \ref{fig:modesDEC}, and \ref{fig:modesSEC} for three energy 
conditions.  Then the reconstruction we need can be expressed as a linear 
combination of the principal components
\begin{eqnarray}
    \label{eq:linearcombination}
    F(z) = \sum_{i} a_i f_i(z).
\end{eqnarray}
Notably, the first three principal components appears stable regardless of 
which energy condition the reconstruction is for.  However, the higher-order 
and noisier modes show greater change in the form with respected to the energy 
condition being reconstructed. Furthermore, when reconstructing $F(z)$ by 
fitting the linear expression (Eq.\ [\ref{eq:linearcombination}]), the 
curvature parameter $\densk$ appearing in equations (\ref{eq:nec}) and 
(\ref{eq:dec}) has been marginalized with numerical integration.

\begin{figure}
    \includegraphics[width=.45\textwidth]{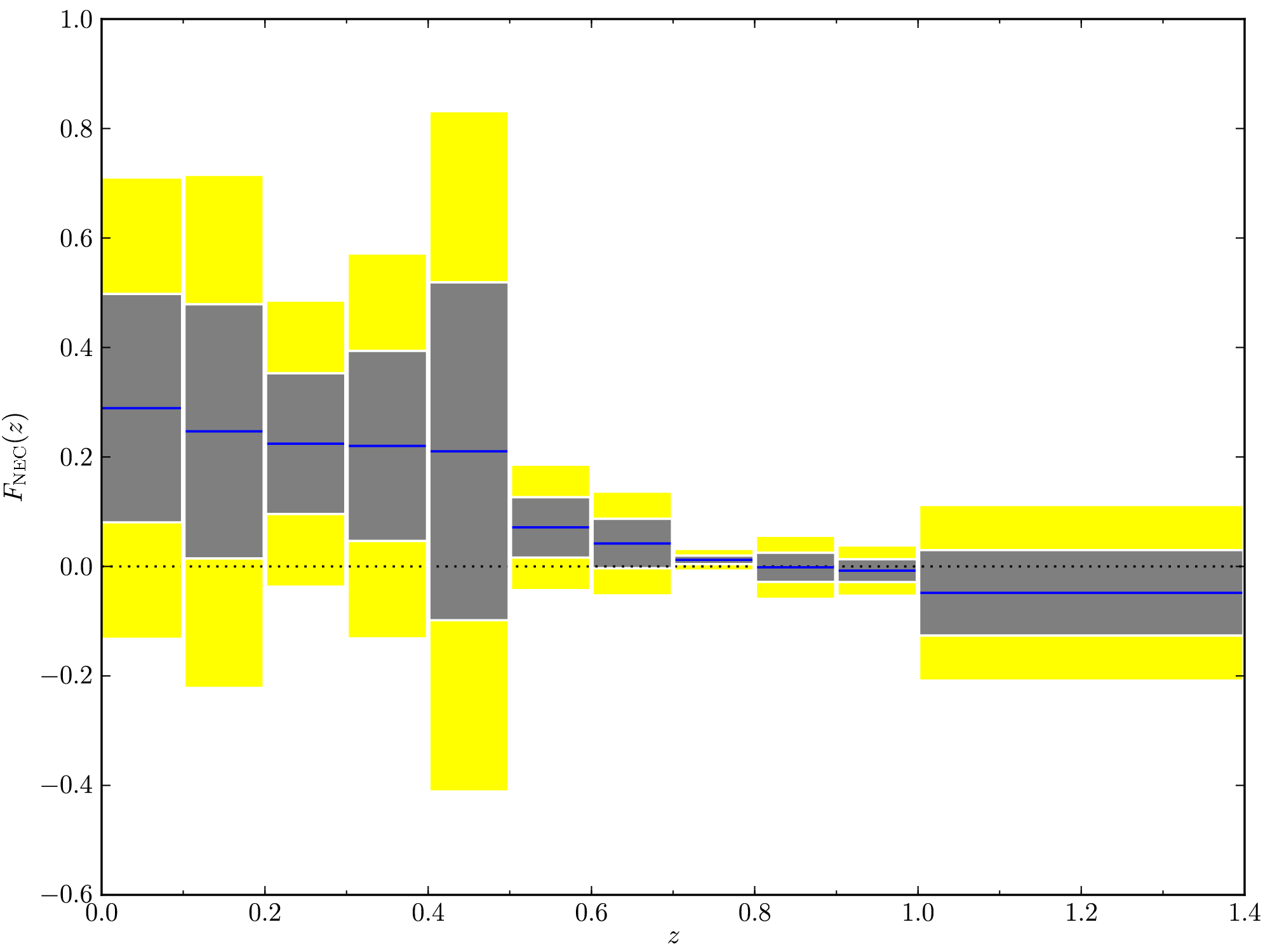}
    \caption{Reconstruction of $F_{\mathrm{NEC}}(z)$ using the first three 
    eigenvectors. The central blue solid line in each bin is the best 
    three-mode fitting result.  The gray and yellow bands show 1- and 
    2-$\sigma$ uncertainties.\label{fig:recNEC}}
\end{figure}

\begin{figure}
    \includegraphics[width=.45\textwidth]{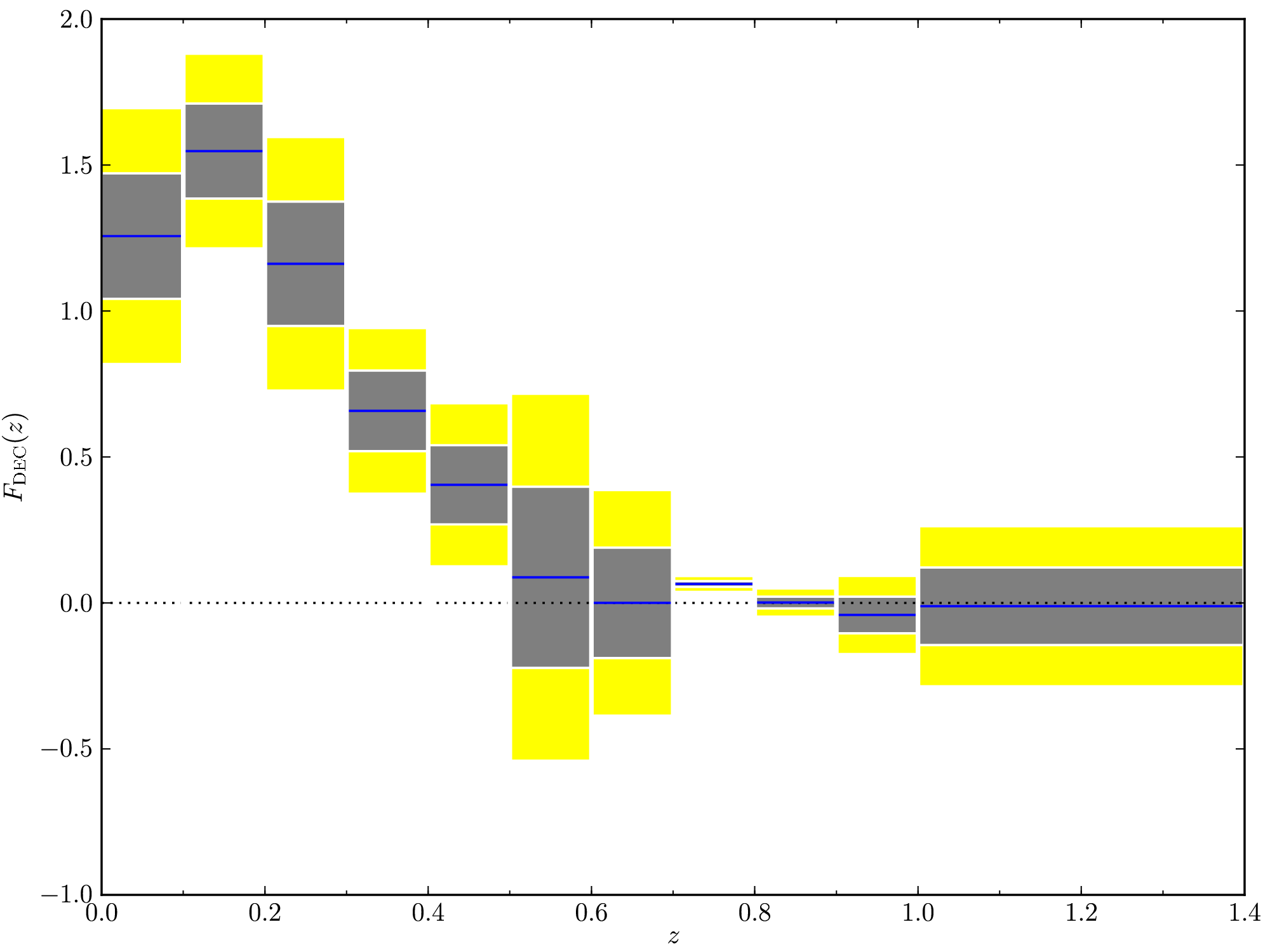}
    \caption{Same as Figure \ref{fig:recNEC} but for 
    $F_{\mathrm{DEC}}(z)$.\label{fig:recDEC}}
\end{figure}

\begin{figure}
    \includegraphics[width=.45\textwidth]{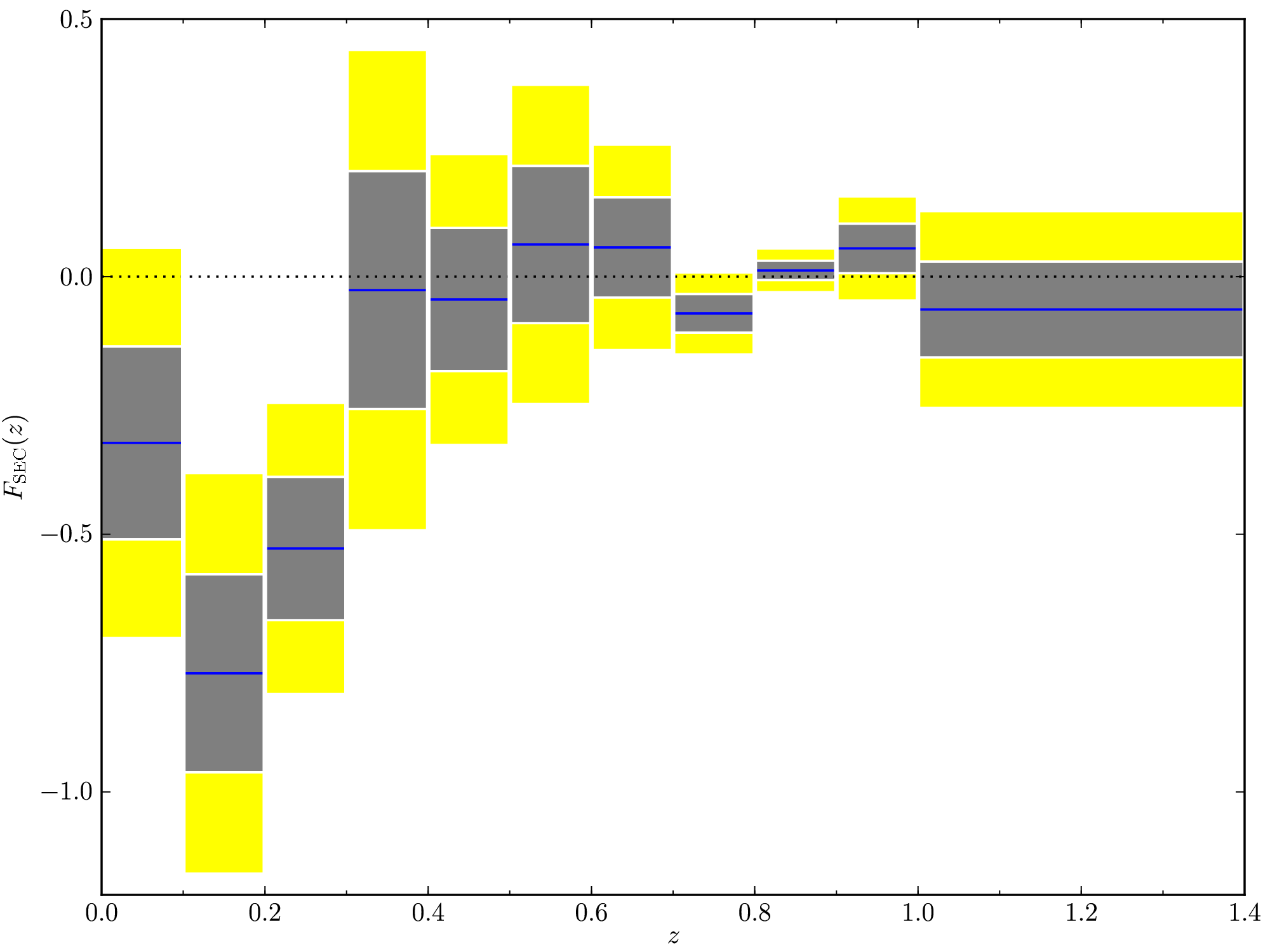}
    \caption{Same as Figure \ref{fig:recNEC} but for 
    $F_{\mathrm{SEC}}(z)$.\label{fig:recSEC}}
\end{figure}

The reconstructed $F(z)$ for the three energy conditions are shown in Figures 
\ref{fig:recNEC}, \ref{fig:recDEC}, and \ref{fig:recSEC}.  We have use only the 
first three principal components in the fitting of $F(z)$.  The fitted values 
of the coefficients are listed in Table \ref{tab:bestfit}.

\begin{deluxetable}{lccc}
    \tablecaption{Best-fit coefficients of the first three
    principal components\label{tab:bestfit}}
    \tablehead{\colhead{$a_i$} & \colhead{$\phm{-}$NEC} & 
    \colhead{$\phm{-}$DEC} & \colhead{$\phm{-}$SEC}}
    \startdata
    $a_1$ & $-0.51\pm0.08$ & $\phm{-}2.40\pm0.10$ & $\phm{-}0.88\pm0.08$\\
    $a_2$ & $\phm{-}0.12\pm0.21$ & $-0.31\pm0.22$ & $-0.18\pm0.21$\\
    $a_3$ & $-0.16\pm0.45$ & $-0.25\pm0.50$ & $-0.43\pm0.40$
    \enddata
\end{deluxetable}

This cutoff from the full spectrum of principal components introduces a 
well-known artifact, namely a bias at the higher end of the redshift 
\citep{2006ApJ...649..563S} that suppresses the reconstructed value and its 
error bars towards zero.  Nevertheless, we note that a value of $F(z)$ close to 
zero is consistent with our {\em lack of knowledge} about the violation of 
energy conditions by the very construction of $F(z)$.  In other words, a 
vanishing value of $F(z)$ tells us that our bet of finding energy condition 
fulfillment is no significantly higher or lower than that of finding its 
violation.  Therefore, this bias in the value of $F(z)$ does not tempt us with 
{\em false claims} about energy condition violations, but merely restates our 
ignorance about it at redshift ranges where there are insufficient data.

\section{Conclusion and Discussions}
\label{s:conc}

We have explored the likelihood of energy condition violation during the 
evolution of the universe using the indication functions we proposed in Section 
\ref{s:ec:overview}.  The indication functions are reconstructed from SNIa and 
$H(z)$ data.

Our result for NEC does not prefer a history of violation, as is evident from 
Figure \ref{fig:recNEC}.  A slight but rising trend can be seen from $z \approx 
0.7$ up to now.  This trend is more pronounced in our result for DEC presented 
in Figure \ref{fig:recDEC}, and we can be fairly certain that there has been no 
DEC violation for $z \le 0.5$.

However, the result for SEC shows fairly strong indication of violation.  This 
shows the view of an accelerating cosmic expansion is compatible with our 
combination of data, especially for low-redshift ($z \le 0.3$).  Although we 
have noted the diminishing ability to distinguish energy condition fulfillment 
from violation at the high-$z$ end, we nevertheless find that our result for 
SEC hints at a recent transition from deceleration ($F_{\mathrm{SEC}}(z) = q(z) 
> 0$) to acceleration ($F_{\mathrm{SEC}}(z) = q(z) < 0$) with the transition 
redshift $z \approx 0.5$, if we ignore the probably biased result at the high 
redshift end.  This estimation of transition redshift is consistent with the 
transition redshifts reported by \citet{2004ApJ...607..665R} and 
\citet{2006ApJ...649..563S} using a kinematic model with $q(z)$ linear in $z$.

For all energy conditions considered, our results show qualitative agreement 
with previous studies \citep{1997Sci...276...88V,1997PhRvD..56.7578V,
2000ppeu.conf...98V,2003A&A...402...53S,2008CQGra..25p5013C}.  In particular, 
the clear trend of DEC fulfillment at low redshift is similar to what was found 
by \citet{2008PhRvD..77h3518L,2008PhLB..668...83L}.

These results are obtained without reference to any specific form of dark 
energy.  With this in mind, we can use these results to assess the feasibility 
of theoretical dark energy models.

Some models predict that the dark energy component undergoes a transition into 
the ``phantom'' phase with the equation of state parameter $w < -1$ 
\citep[][and references therein]{2005PhRvD..71l4036S}.  It has been noted that 
NEC (hence necessarily DEC) violation is a feature of these models, and some 
oscillating universe models predict episodes of NEC violation throughout the 
history \citep{2010PhR...493....1C}.  Even if certain phenomenological models 
could offer a fairly good fit \citep{2005MPLA...20.2409X}, our data-driven 
result for NEC suggests that there is not enough motivation to consider them as 
a viable explanation for cosmic acceleration, at least not in the redshift 
range investigated in this paper.

On the other hand, a family of dark energy models exist that do not violate NEC 
or DEC, yet display transient periods of SEC violation 
\citep{2000MNRAS.316L..41B,2004PhRvD..70h4008B,2006JCAP...04..008B}.  While 
such scenarios are unlikely to have been realized from the early universe up to 
$z \gtrsim 1$ \citep{2010PhRvD..82f3514L,2011JCAP...04..001L}, from an 
energy-condition point of view they are not ruled out by our independent 
results in low redshift.  Whether the ``returning'' $F_{\mathrm{SEC}}$ curve in 
Figure \ref{fig:recSEC} within $z < 0.4$ suggests a coming back to SEC 
fulfillment, which has been independently noted earlier 
\citep{2009PhRvD..80j1301S,2011CQGra..28l5026G} for spatially flat models, may 
be worth examining using future data.

As we have already noted in Section \ref{s:results}, our test is sensitive in 
low redshift but not in high redshift.  In the future, this could be partially 
alleviated by the availability of high-quality cosmological data that could 
populate higher redshift.

Nevertheless, our work addresses several deficiencies in previous literature 
and has some special merits.

First, by proposing a set of new models based on our indication functions, we 
are able to visualize the evolution of the Universe as a history of energy 
condition violation or fulfillment in a straightforward manner.  By 
construction, we can avoid ambiguous statements such as ``SEC implies $H_0 (z_f 
\approx 15) \le 58 \pm 7$ $\hunit$'' \citep[paraphrasing Eq.\ 
\protect{[}79\protect{]} of][]{1997PhRvD..56.7578V}, where the inequality 
nature of the energy conditions find some difficulty being used with error 
bounds from observational data.  Compared with the research done by 
\citet{2008PhRvD..77h3518L}, our models and their presentations can be more 
readily interpreted as answers of ``whether an energy condition was violated 
for a given epoch'', because our indication functions stay constant in each 
redshift bin.

Second, our models are general and are free from overly restrictive 
presumptions on the nature of the dark energy.  It stands in contrast to 
previous studies by \citet{2003A&A...402...53S} where a constant equation of 
state parameter model was used to study the possibility of energy condition 
violation.

Third, by incorporating the Fisher matrix formalism in our analysis, we find an 
optimal set of bases for the analysis of energy condition violation in the 
sense that the coefficients for different modes are nearly uncorrelated.  When 
we generate this set of bases, the effects of $\densk$ are naturally and 
explicitly taken into account.  Compared with the parametrization of 
\citet{2006ApJ...649..563S}, our method is more generally applicable because we 
do not presume spatial flatness, and has a unified description for all the 
energy conditions considered.  We apply this formalism to the combination of 
SNIa and $H(z)$ data, and this can be extended to other suitable cosmological 
probes when the data become available.

\begin{acknowledgments}
C.-J.\ Wu is grateful to Liang Gao, Hong Wu, and Ming-Jian Zhang for their kind 
help.  C.\ Ma thanks Hao Wang and Kai Liao for valuable discussions.  This work 
was supported by the National Science Foundation of China (Grants No.\ 
11173006), the Ministry of Science and Technology National Basic Science 
program (project 973) under grant No.\ 2012CB821804, and the Fundamental 
Research Funds for the Central Universities.
\end{acknowledgments}

\bibliographystyle{hapj}
\bibliography{ms}
\end{document}